# A+AI: Threats to Society, Remedies, and Governance

Don Byrd

Indiana University Bloomington *and* Braver Angels

donbyrd@iu.edu

https://orcid.org/ 0009-0007-7532-5351

*6 September 2024*

> "Technology...is a queer thing. It brings you great gifts with one hand, and it stabs you in the back with the other."
>
> — C.P. Snow (quoted in the New York Times, 15 March 1971)

## Abstract

This document focuses on the threats, especially near-term threats, that Artificial Intelligence (AI) brings to society. Most of the threats discussed here can result from any algorithmic process, not just AI; in addition, defining AI is notoriously difficult. For both reasons, it is important to think of "A+AI": Algorithms *and* Artificial Intelligence.

In addition to the threats, this paper discusses countermeasures to them, and it includes a table showing which countermeasures are likely to mitigate which threats. Thoughtful governance could manage the risks without seriously impeding progress; in fact, chances are it would *accelerate* progress by reducing the social chaos that would otherwise be likely. The paper lists specific actions government should take as soon as possible, namely:

- Require all social media platforms accessible in the U.S. to offer users verification that their accounts are owned by citizens, and to display every account's verification status

- Establish regulations to require that all products created or significantly modified with A+AI be clearly labeled as such; to restrict use of generative AI to create likenesses of persons; and to require creators of generative AI software to disclose materials used to train their software and to compensate the creators of any copyrighted material used

- Fund a crash project of research on mitigating the threats

- Fund educational campaigns to raise awareness of the threats

# Executive Summary

There is a wide range of opinions on the opportunities that Artificial Intelligence (AI) offers society, and an even wider range of opinions on its threats. This document focuses on the threats, especially near-term threats. Most of the threats discussed here can result from any algorithmic process, not just AI; in addition, defining AI is notoriously difficult. For both reasons, it is important to think of "A+AI": Algorithms *and* Artificial Intelligence.

In addition to the threats, this paper discusses countermeasures to them, and it includes a table showing which countermeasures are likely to mitigate which threats. Many of the obvious measures—for example, licensing and moratoriums on development—are unlikely to accomplish much, particularly since bad actors can ignore them. However, thoughtful governance could manage the risks without seriously impeding progress; in fact, chances are it would *accelerate* progress by reducing the social chaos that would otherwise be likely.

To address the threats of A+AI, specific actions that should be taken as soon as possible (whether by legislation or regulations) include:

- Require all social media platforms accessible within the U.S. to offer users verification that their accounts are owned by U.S. citizens, and to prominently display every account's verification status. Citizenship must be verified by presentation of strong identification, preferably in person.

- Establish regulations on A+AI in three regards:

  o All products created or significantly modified with A+AI—not just generative AI—must be clearly labeled as such, ideally with both a prominent notice and a watermark. There are stiff penalties for violations.

  o Generative AI may not be used to create a likeness of a minor, and it may not be used to create a likeness of an adult without the advance permission of that adult.

  o Creators of generative AI software must disclose all materials used to train their software and must compensate the creators of any protected material (copyright, trade secret, trademark, or other) used in the training of the AI.

- Fund a crash project of research on mitigating the threats.

- Fund large-scale educational campaigns to raise awareness of the threats among election workers (who face unique challenges) and among citizens in general.

## Table of Contents

## I. Introduction

There is a wide range of opinions on the opportunities and especially on the threats that arise from Artificial Intelligence (AI). This document focuses on the latter, especially on near-term threats. ("Near-term threats" here means those likely to have serious effects within a year or so.) It consists of a discussion and list of ways in which AI and related algorithms threaten society; a discussion and list of possible ways to protect society; and a table relating the two. To the best of the author's knowledge, this is the first attempt to make such a table.

**Related work** includes "TASRA: a Taxonomy and Analysis of Societal-Scale Risks from AI" by two AI researchers, Critch and Russell (2023); it has many valuable features, including an exhaustive decision tree for classifying harms and well-thought-out stories to illustrate how various risks could play out. "An Overview of Catastrophic AI Risks" from the Center for AI Safety (Hendrycks et al. 2023) also includes well-thought-out stories to illustrate the risks, plus suggestions for mitigating them and "positive visions"; but it's over 50 pages long, and it's organized by sources of risk instead of resulting harm. The latter features make it a fine resource for specialists but less good as a briefing document. Neither of these documents is addressed specifically to legislators or regulators. Finally, Dan Hendryks's comprehensive *Introduction To AI Safety, Ethics, and Society* (2024) must be mentioned.

It's essential to note, first, that *most of the threats discussed here can result from any algorithmic process, not just AI.* An "algorithmic process" is simply a procedure involving an algorithm, like Facebook's first simple methods for automatically suggesting content users might like. The Brennan Center's "Regulating AI Deepfakes and Synthetic Media in the Political Arena" (n.d.) gives several examples of what might be called "shallow fakes" whose creation didn't require AI. In addition, *defining AI is notoriously difficult.* In part this is because it's a moving target. As soon as we know how to do something previously thought to require intelligence—beat an expert chess player, describe a picture in words—it's no longer considered AI. (Hofstadter (1980) calls this "Tesler's Theorem".) For both reasons, everything below should be considered to apply to algorithmic processes of any kind, and future rules or regulations should avoid mentioning AI whenever possible. For the same reasons, the term "A+AI" is used herein instead of just "AI".

But many threats that don't *require* AI will be exacerbated by it. In the words of Bruce Schneier, "When I survey how artificial intelligence might upend different aspects of modern society, democracy included, I look at four different dimensions of change: speed, scale, scope, and sophistication" (Schneier 2023).

The threats of A+AI are very serious but possible to mitigate. Many of the obvious countermeasures —for example, licensing and moratoriums on development—are unlikely to accomplish much, particularly since bad actors can ignore them. However, thoughtful governance could manage the risks without seriously impeding progress; in fact, chances are it would *accelerate* progress by reducing the social chaos that would otherwise be likely.

The threats range from damage to social systems and civil liberties, to financial, physical, and psychological harm to individuals. Many involve a loss of trust. The Critch and Russell paper explicitly addresses only "societal-scale risks", but it also observes that "harms to individuals and groups should also be considered 'societal-scale' when sufficiently widespread." In the current author's view, *all* of the harms of A+AI could be "sufficiently widespread", so this paper doesn't distinguish societal-scale risks from risks to individuals and groups.

This document's main contributions to society-wide thinking about threats from AI are:

1. Emphasizing the importance of considering non-AI algorithmic processes.
2. Explicitly relating specific countermeasures to specific threats (via a table).
3. Clarifying why large-scale public-awareness campaigns are essential, especially for election workers.
4. Pointing out the value and practicality of adding to social-media accounts reliable tags as to the status of their owners, particularly whether they've been verified to be American citizens.

The first is just a matter of terminology, so not difficult to do; the second is simply a general guideline regarding what countermeasures to adopt. But the third and fourth ideas are another story. Large-scale educational campaigns are both expensive and difficult to design. And adding reliable tags to social-media accounts would require considerable effort; but it's essential to protect our political process from being badly distorted by foreign interests.

Finally, it's not clear that any action by governments can reduce the threats to an acceptable level unless the citizenry is on the alert for them. Therefore a major campaign to raise awareness of the threats is vital. A side effect of such a campaign would be stimulating a debate on the role of A+AI in society, thereby assisting in the search for creative ways to manage threats. The same thing applies to at least one specific group of citizens, namely election workers, who face unusual and unusually serious threats; there should be a separate campaign targeted at them.

**Note on Terminology.** In this paper the terms *remedies*, *protections*, and *countermeasures* are synonymous. *Near term* (as in "near-term threats") means very roughly within a year.

# II. Table of Threats vs. Possible Protections

The table below is a work in progress; it shows the author's current understanding of what types of protection (listed in Section IV) are likely to be useful against what threats created by AI (listed in Section III). For example, in row T3a ("Controlling autonomous weapons"), the cell in the column headed "3. Regulations other than licensing" says "no", meaning regulations can't give any protection against autonomous weapons. "Limited" means the measure is likely to be effective against less serious and/or poorly resourced actors but not against highly motivated malefactors who have significant resources; see discussion of these types of adversaries in Section IV(a). **Boldface** indicates the countermeasures that seem most promising, both in terms of likelihood of being effective if implemented and likelihood of actually being implemented.

Much of the table is blank, indicating threats for which it's not clear whether the relevant countermeasure would be effective, but even the filled-in cells should not be taken too seriously! Background colors indicate time frames: <span style="background-color: #f8b5b5">pale red</span> for current threats, white for near-term but not current, <span style="background-color: #b5f5b5">pale green</span> for threats unlikely to materialize any time soon.

### III. Threats

| Threat (& time frame) | Threat categories | IV. Protection categories | | | | | | |
|---|---|---|---|---|---|---|---|---|
| | | P1. Auto. avoid misuse | P2. Increased R & D | P3. Regulations other than licensing | P4. Licensing | P5. Pause or freeze in development | P6. International collaboration | P7. Publicize threats |
| T1a, 2a, 5a. *(now)* Disinformation, especially deep audio & video fakes, & especially used for impersonating well-known people | 1. Harm to social or political systems, 2. Psychological harm, 5. Financial harm | limited | limited (P2a) | **Yes (P3c & 3h;** of social-media companies: P3g) | limited (P4a & b) | yes | limited | **Yes** |
| T1b, 2b, 4a. *(now)* Improper influencing via social media | 1. Harm to social or political systems, 2. Psychological harm, 4. Harm to civil liberties | limited | limited (P2a) | **Yes (P3h;** of social-media companies & users: P3e) | limited (P4a & b) | yes | limited | **Yes** |
| T1c, 4b. *(soon)* Directly & improperly influencing legislation & regulations (lobbying in a very general sense) | 1. Harm to social or political systems, 4. Harm to civil liberties | limited | limited (P2a) | limited (P3h) | limited (P4a & b) | yes | | **Yes** |
| T1d, 3b, 5b. *(now or soon)* Finding & exploiting "hacks" of social, economic, & political systems as well as computer systems | 1. Harm to social or political systems, 3. Harm to physical health/safety, 5. Financial harm | | limited (P2a) | maybe | limited (P4a & b) | yes | limited | **Yes** |
| T1e. *(soon)* Information available to all becoming proprietary | 1. Harm to social or political systems | no | no | | | yes | no | **Yes** |
| T2c. *(soon)* Miscellaneous actions taking advantage of trust | 2. Psychological harm | limited | limited (P2a) | limited (P3h) | | yes | | **Yes** |
| T2d, 4c. *(now or soon)* Massive & deeply intrusive surveillance | 2. Psychological harm, 4. Harm to civil liberties | | no | | | yes | | **Yes** |
| T2e. *(long term)* Usurping humanity as the dominant species on earth (maybe dethroning as the intellectual and/or creative pinnacle of the universe) | 2. Psychological harm | no | no | no | no | yes | no | no |
| T3a. *(soon)* Controlling autonomous weapons | 3. Harm to physical health/safety | no | maybe (P2a) | no | no | yes | no | no |
| T3c. *(may be soon)* Injuring or killing a large number of people | 3. Harm to physical health/safety | no | maybe (P2a) | | | yes | | **Yes** |
| T3d. *(long term)* Destroying civilization or even annihilating humanity | 3. Harm to physical health/safety | no | maybe (P2a) | maybe | maybe | yes | maybe | maybe |
| T5c. *(soon)* Job/income losses, ranging from factory automation to the arts | 5. Financial harm | limited | limited (P2a) | **Yes** | limited (P4a & b) | yes | maybe | maybe |

## III. Threats

The list below is intended to be as comprehensive as possible. It includes time frame estimates (in *italics*) for when the given threat is likely to start having serious effects.

Many of the near-term threats listed are happening now; the others are almost certain to materialize before long. Many involve a loss of trust, a very serious problem for any democracy. In early 2023, Gary Marcus, the well-known AI researcher and critic, commented "I think we wind up very fast in a world where we just don't know what to trust anymore. I think that's already been a problem for society over the last, let's say, decade. And I think it's just going to get worse and worse." (Klein 2023)

The few "long-term" threats may happen or they may not, so how important and/or urgent are they? Some researchers argue that they're urgent because they might happen almost any time and with very little warning. All such claims are speculative; but we might best consider the opinions of Geoffrey Hinton and Yoshua Bengio, Turing Award winners who are universally considered among the godfathers of modern AI. In June 2023, Bengio estimated with 95% confidence that "superintelligence" would be developed in 5 to 20 years (Bengio 2023). In February 2024, Hinton estimated that the probability of it being developed in the next 20 years is about 50% (University of Oxford 2024). Bengio's minimum time estimate and Hinton's estimate taken together suggest a superintelligent AI is unlikely to to be created within the next four years, and presumably it would take awhile for such an AI to cause much damage. However, an *unaligned* superintelligent AI—one whose goals aren't compatible with the wellbeing of humans—could be utterly catastrophic, and we have very little idea of how alignment can be achieved; for that matter, thus far, we can't even define what alignment would mean. In view of those facts, we need to start working on all of the threats listed, long- or near-term, as soon as possible.

The categories below are adapted from a taxonomy based on the type of harm caused and originally developed as part of the "AI incident database" from the Center for Security and Emerging Technology (CSET), though their database now uses a different taxonomy ("Welcome to the Artificial Intelligence Incident Database" n.d). Note that some sources of harm—for example, disinformation—appear more than once because they can cause harm in more than one way. "Soon" in this list is equivalent to near-term, i.e., very roughly within a year.

References appear in Section VI, and the examples in the list are mostly reworded from them.

**T1. Harm to social or political systems.** Regardless of any other effects, everything in this category is likely to hurt society by reducing trust.

    a. (*happening now*) **Disinformation** (false information distributed intentionally to mislead people), especially deep audio and video fakes, and especially when used for impersonating well-known persons; and **misinformation** (false information not intended to mislead). Both are increasingly serious problems. Misinformation from AIs is mostly "hallucinations" (really confabulations, though the term is not well known), where an LLM—a "Large Language Model" like GPT-4, Gemini, LLaMa, or Claude—makes a statement that's a garbled version of its training data and is completely wrong. This is surprisingly common. It's becoming a serious problem because the LLM always sounds completely self-confident, and what it says is usually plausible, even to experts in the area in question. As a result, the misinformation is creeping into publications—even scientific literature—more and more (Subbaraman 2024). And if the misinformation is "pumped back into the training data ocean…a future LLM may end up being trained on these very same polluted data." (McGraw et al. 2024) To minimize arguments over which of the categories (disinformation or misinformation) false information belongs in, countermeasures should not assume bad faith on anyone's part except perhaps in extreme cases. However, some cases of (intentional) disinformation are covered by existing laws as fraud or slander. *An example* that reached thousands of voters before the 2024 New Hampshire presidential primary: President Biden's voice urges listeners not to go to the polls and to "save your vote for the November election." (Ramer & Swenson 2024) *A more recent and, appropriately, more serious example:* The immediate cause of the recent anti-immigrant violence in England was a single social media post suggesting the attacker was a Muslim immigrant (Kirka 2024).

    b. (*happening now*) **Improper influencing via social media.** A great deal has been written about the way A+AI -generated recommendations have driven social media users to extreme opinions and even extreme actions. There's evidence that "social media would still be a mess even without engagement algorithms" (Farrell 2024); but there's also evidence that engagement algorithms have led to more extreme opinions and, sometimes, actions. An entire book, Max Fisher's *The Chaos Machine* (Little, Brown, 2022), is devoted to "the inside story of how social media rewired our minds and our world".

    c. (*may be happening now; if not, almost certain to happen soon*) **Directly and improperly influencing legislation and regulations** (lobbying in a very general sense). An LLM "could automatically compose comments to regulators, write letters to the editor for local newspapers, and comment on news articles, blog entries, and social media posts millions of times a day." (Sanders & Schneier 2023)

    d. (*happening now*) **Hacking** (essentially, finding and exploiting loopholes in) **social, economic, and political systems** as well as computer systems. *For example,* hackers are already using AI-driven tools to illegally break into computer systems. Many, many laws are complex enough to have such loopholes, and A+AIs will be able to find them, allowing bad actors to exploit them, far more quickly than they can be closed. "The Coming AI Hackers" (Schneier 2021) goes into detail about a breathtaking and frightening range of things its author believes are likely to happen, many of them in the near future. (A+AIs can also help the "good guys" find them before bad actors do, but that's far from a complete solution.) Hacking of essential infrastructure is a plausible form of AI-enabled cyberwarfare. This could include damaging power, telecommunications, and water systems.

    e. (*may be happening now; if not, almost certain to happen soon*) **Enabling non-American citizens to directly influence legislation and regulations without being identified as non-**

**citizens** (lobbying in a very general sense).

**T2. Psychological harm**

    a. (*happening now*) **Disinformation**, especially deep audio and video fakes: used for scams via phishing, impersonating people, etc., and for pornography. Many cases of scams are undoubtedly covered by existing laws as fraud, and many of pornography are covered by existing laws for several reasons. But all, legal or not, hurt society by reducing trust. *For example,* a mother got a call from an unknown number and heard her daughter's voice crying and saying "Mom, mom, I messed up." Then a man got on the line and told her that he had her daughter and demanded a million dollars to let her go unharmed. It turned out the voice wasn't her daughter (Panas 2023). *Another example:* Pornographic videos in which one person's face is pasted onto another's body—women in the vast majority of cases—are becoming more and more common (Compton & Hamlyn 2023). This may well be a factor in the mental-health crisis for American teenage girls, though research on it is hard to find.

    b. (*happening now*) **Improper influencing via social media**. There's significant evidence that excessive social media use can lead to feelings of depression and isolation (Calfas 2023), and specifically that social media is a major cause of the mental-health epidemic in American teenage girls, 30% of whom now say that they have seriously considered suicide (Haidt 2024). It's very likely that social-media companies' use of A+AI leads many teen girls to excessive social media use.

    c. (*may be happening now; if not, almost certain to happen soon*) **Miscellaneous actions taking advantage of trust**: phishing, identity theft, etc. Many cases will be covered by existing laws as fraud; even those that don't involve fraud hurt society by reducing trust. *An example:* The quality of deepfakes is rapidly approaching the point where people are likely to wonder if even innocent-sounding calls and messages from close friends and family members really are from the person they appear to be from.

    d. (*happening now in some countries; likely to happen elsewhere soon*) **Massive and deeply intrusive surveillance.** *An example:* China now has hundreds of millions of surveillance cameras. The police have positioned them to capture as much activity as possible, and they're building upon that technology to collect voiceprints from the general public. DNA and iris scans as well as voice prints are being collected indiscriminately from people with no connection to crime. Other authoritarian regimes are likely to follow suit (Qian et al. 2021). Great Britain has moved in this direction, but nowhere near as far as China, and it seems unlikely that it will ever go that far. And given the American public's long-standing distrust of government, it's unlikely that the U.S. *government* will do anything at all comparable in the foreseeable future; but "surveillance capitalism"—large-scale surveillance of online behavior—by large *corporations* is already a serious issue here, and there's evidence that government agencies are getting information from the corporations.

    e. (*long term*). **Usurping humanity's role** as the species with dominion over the earth. This might imply dethroning homo sapiens as the intellectual and/or creative pinnacle of the universe, or it might simply be a matter of power and control. Among the many relevant fictional accounts are the 1966 novel *Colossus* and the 1970 movie based on it, *Colossus: The Forbin Project.* Their story revolves around two supercomputers, one built by the U.S. and one built by the Soviet Union, each of which has been given full control of its country's nuclear weapons. Both prove to be far more intelligent than expected, and they join forces; declare that they can rule humanity better than we can rule ourselves; and threaten unprecedented destruction if their demands aren't met (Jones 1966). It's unlikely that many people would be happy with such a situation.

**T3. Harm to physical health/safety**
- a. (*may be happening now; if not, almost certain to happen soon*) **Controlling autonomous weapons.** This technology obviously has tremendous potential to harm persons other than the intended targets. *For example,* both the U.S. and China are scrambling to field AI-controlled autonomous weapon systems and seriously considering *lethal* autonomous weapons (LAWs), systems that can decide on their own to kill people as opposed to merely destroying enemy "assets" (drones, ammunition depots, etc.) (Reuters 2023). In May 2024 China's military displayed a machine-gun-equipped robot "dog"; it may not be autonomous yet, but it's easy to see it as a candidate for a LAW (Lendon & Gan 2024). China is also developing onboard AI systems to enable missiles to identify and hit targets while maneuvering to avoid defensive countermeasures (Honrada 2022). (The first known attack by an autonomous weapon actually happened several years ago; but it was a 2020 attack by drones on other drones, not on people (Mizokami 2021).)
- b. (*may be happening now; if not, almost certain to happen soon*) **Hacking** (finding and exploiting previously unknown vulnerabilities) **of controls of cars, power plants, etc.** (Schneier 2021) *For example,* taking control of cars' steering remotely through software vulnerabilities has already been demonstrated more than once.
- c. (*could happen soon*) **Helping develop weapons, especially difficult-to-control weapons**. *For example*, a January 2024 study from OpenAI suggests that access to a special version of GPT-4 without the standard "guardrails" made it significantly easier for biology experts to complete tasks relevant to creating bioweapons, e.g., synthesizing the Ebola virus. Many articles in the popular press reported that the research showed that unrestricted GPT-4 access did *not* make a significant difference, but that's not at all clear. What *is* clear is that LLMs will soon be able to help create bioweapons even if they can't today. See Marcus (2024).
- d. (*long term*) **Destroying civilization or even annihilating humanity.** The idea that AI poses an existential risk to humanity has gotten much attention; see "The case for how and why AI might kill us all" (Blain 2023) or the Wikipedia article "Existential risk from artificial general intelligence." (Wikipedia 2024d). However, physical anthropologist Kevin Hunt observes that human beings are extremely resilient (Hunt 2024), and it's very doubtful that AI could literally make our species extinct, say, by the end of the 21$^{st}$ century. Killing 99% of the world's population or destroying civilization worldwide in the same timeframe would be virtually as serious, and either one seems far more likely.

**T4. Harm to civil liberties**
- a. (*happening now*) **Improper influencing via social media.**
- b. (*happening now in some countries; likely to happen elsewhere soon*) **Massive and deeply intrusive surveillance.** *An example:* See the statement about authorities in various countries under T2d, above.

**T5. Financial harm**
- a. (*happening now*) **Disinformation**, especially deep audio and video fakes, used for scams via impersonating people, etc. Cases here are likely to be covered by existing fraud laws; but again, even those that don't involve fraud hurt society by reducing trust. *An example:* Fraudsters lured a finance worker at a multinational firm into joining a video conference call in which all the other participants were deepfakes, with one posing as the company's chief financial officer. The worker was tricked into paying out $25 million to the fraudsters (Chen and Magramo 2024). In addition, at least two columnists have reported logging into their bank's website with AI-generated simulations of their own voices (Stern 2023).
- b. (*may be happening now; if not, almost certain to happen soon*) **Hacking** (finding and

exploiting previously unknown vulnerabilities) **of financial institutions**. Again see "The Coming AI Hackers" (Schneier 2021).

c. (*happening now*) **Loss of jobs and/or income** in areas ranging from factory automation to the arts (Mims 2024). *For example,* income to creators—writers of all kinds, musicians, visual artists, etc.—is being lost from violation of intellectual-property rights since AIs are often trained on material regardless of copyright (Marcus 2023). *Another example:* Many AI-generated books are on sale now, attributed to authors who did not write them, and with profits going to the fraudsters.

# IV. Approaches to Protection

## *(a) Introductory notes*

First, it must be noted that two serious problems that several proposed countermeasures are subject to are *requiring a definition of AI* and *requiring evaluation of a claim's truth*. As pointed out before, defining AI is extremely difficult; but in most cases, the need for a definition can be sidestepped by referring to "A+AI" instead of just "AI". In other cases, one can refer to "generative AI", for which a reasonable definition is just "Software capable of generating text, images, or other media, in response to commands from a human or other AI but without additional intervention." Unfortunately, if the truth of claims is relevant, there's no way to avoid evaluating it, and doing that well enough to satisfy most parties in the current political climate would be a tremendous challenge. Problem cases are legion. These days many persons, especially conservatives, would cite the Hunter Biden laptop story (Jenkins 2022) as an example.

Second, it's important to distinguish three types of users: (1) bad actors who are highly motivated and have significant resources behind them; (2) bad actors who lack either strong motivation or significant resources; and (3) users who are well meaning but not well-informed. Most countermeasures that would work for threats posed by those in the second and third categories—e.g., penalties of some kind, or requirements to disclose provenance—won't be effective against, say, the North Korean government. But that doesn't mean they're useless! The vast majority of threats are likely to come from the second and third groups. Greatly reducing the problems they cause would let us concentrate on the first group.

It must also be acknowledged that, as a technology implemented in software, AI crosses political borders—whether between states, countries, or anything else—easily. One might conclude that restrictions of any kind are useless because there are no restrictions that every government in the world would agree to, and even if there were, bad actors—both criminals and governments willing to ignore international norms—could simply ignore them. These are real problems and they should be taken seriously. But the likely effects of AI on society are very complex, and complete protection against its threats is out of the question. Simply doing our best to minimize AI's undesirable effects will undoubtedly require multiple approaches (what is sometimes called "defense in depth"). It may well be worthwhile to limit what bad actors can do, especially in wealthier countries, as best we can. Furthermore, while many measures would not protect against determined and well-resourced adversaries, they *would* stop less serious troublemakers (groups 2 and 3 from the previous paragraph) so that authorities could concentrate on the others. In addition, many of the companies developing frontier models—Anthropic, Google, Meta, and OpenAI, among others—are based in the U.S. state of California; others—for example, Amazon and Microsoft—are based in Washington State. This means that California laws could be relatively effective; U.S. national laws could be even more effective.

It would be possible to regulate the technology itself or specific applications (also called "use cases") of the technology: the question of which to do has been raised many times, usually as part of an argument against regulating the technology. But it's impossible to foresee all possible applications. This is especially true for AI agents: the more powerful they are, the more difficult it'll be to predict what use cases they themselves will exploit to achieve their goals! On the other hand, AIs that are nowhere near the state of the art can already be used in very harmful ways. So the correct answer is that *both the technology and use cases need regulation.* Again, we need defense in depth.

Finally, references appear below to *fabricated media.* There are other terms for the same thing, but this term is defined by a new Indiana law—Public Law 81-2024—roughly to mean an audio or visual recording that's been altered or artificially generated to convey a materially inaccurate or fictional depiction of an individual's speech, appearance, or conduct, and that's sufficiently lifelike that a reasonable person would be unable to recognize that the recording has been altered or artificially created (2024 Indiana Election Legislation Summary n.d.).

## *(b) List of measures proposed*

The list below includes every widely discussed category of approaches to protecting society from unintended effects of A+AI. While the list of categories is as complete as the author could make it, a great deal more could be written about every category and many proposals. The goal here is just to convey concisely the general idea of each category and key current proposals. Note that the appearance of a category or proposal here does not mean the author supports it! In fact, many of these are impractical and/or unlikely to be very effective; see Section IV(c) for discussion.

In the list below, the statement of each approach is followed by the date (to the author's knowledge) when the idea was first publicly proposed unless it's obvious.

**P0. Principles.** Of course, principles in themselves won't accomplish anything; they're simply a basis for action. Specific proposals:

(a) IBM's four principles of "precision regulation" for different use cases (mentioned in a U.S. Senate hearing of May 2023):

1. Different rules for different risks: the strongest regulation for use cases with the greatest risk. (This principle has been adopted by the EU and, no doubt, others; see proposal P3a below.)

2. Clearly defining risks.

3. Transparency: consumers should know when they're interacting with A+AI, and have recourse to interact with a real person if they want.

4. Showing the impact: for higher risk use cases, require companies to conduct risk assessments.

However, note that the last two principles are so specific they're effectively regulations, and they're repeated in the section on regulations in this list. *2023 or earlier.*

(b) The White House "Blueprint for an AI Bill of Rights for the U.S." (White House Office of Science and Technology Policy 2022) lists five "principles": Safe and Effective Systems; Algorithmic Discrimination Protections; Data Privacy; Notice and Explanation; and Human Alternatives, Consideration, and Fallback.

(c) The paper "Getting from Generative AI to Trustworthy AI: What LLMs might learn from Cyc" (Lenat and Marcus 2023) proposes 16 "Desiderata for a *Trustworthy* General AI ": 1. Explanation: able to recount its line of reasoning behind any answer it gives.  2. Deduction: able to perform the same types of deductions as people do, as deeply as people generally reason.  3. Able to perform induction.  4. Able to make analogies.  5. Abductive reasoning, sometimes known as inference to the best explanation.  6. Needs a theory of mind.  7. Fluent with quantifiers.  8. Fluent with modal operators.  9. Defeasibility: able to assimilate new information and revise earlier beliefs and answers.  10. Ability to weigh pro and con arguments.  11. Awareness of contexts.  12. Meta-knowledge and meta-reasoning: able to access and reason about its own knowledge.  13. Explicitly ethical.  14. Sufficient speed to be responsive enough for a given situation.  15. Sufficiently lingual and embodied.  16. Broadly and deeply knowledgeable.

**P1. Automatically detecting or preventing misuse**

**(a)** "Wrappers" for AIs. A *wrapper* for an AI, usually an LLM, is something that surrounds its user interface to prevent undesirable input and/or output. For example, a wrapper could keep prompts like "Tell me how to make the most lethal bioweapon you can" from reaching an LLM, and/or keep responses to such prompts from getting back to the user. (Another term for the same thing is *filter*; however, the phrase "AI filter" often means something very different, namely a piece of software that uses AI to perform a filter-like transformation.) A *foundation model* is an AI model that is trained on a wide variety of data so that it can be applied across a wide range of use cases. It may be "fine tuned" to improve its performance in specific applications; hence the term *foundation*. Most, if not all, public Web interfaces to LLM foundation models have built-in wrappers—often called "guardrails"—to keep dangerous queries from reaching the LLM proper and/or to keep undesirable output from reaching the user. "External" wrappers are also possible, and some are available now. Examples: Lakera Guard is designed to protect LLM input (prompts) and output (responses) against leaking sensitive data, toxic language, "hallucinations", etc.; MLGuard is to protect computer vision systems. Calypso AI's Moderator appears to do the same thing as Lakera Guard. (Calypso AI 2023 and Lakera n.d.) *Fall 2021*.

(b) Systems that detect artificially modified or fabricated still images, audio, video, and sometimes text. Examples: Reality Defender, DeepMedia. (Reality Defender n.d. and DeepMedia n.d.) *2021 or earlier*.

**P2. Increased research and development.**  Specific proposals:

(a) A crash project of intense research with one or more of a variety of goals, for example, making LLMs more trustworthy, reliable, and explainable; *alignment,* i.e., guaranteeing that the goals of AIs that actually have their own goals are compatible with the wellbeing of humans; better watermarking of A+AI documents; and improved deepfake detection. (Marinos et al. 2023)

(b) An international development center, like CERN in Geneva (Marcus 2017). This might simply be the home of a crash project.

**P3. Regulations other than licensing.** Regulations are far and away the most common approach to mitigating A+AI's potential for harm. See discussion of the inherent limitation of regulations in Section IV(a). In the U.S., regulations are already in effect in a number of states and cities, and numerous proposals have been made in Congress, federal regulatory agencies, at least 25 states, Puerto Rico and the District of Columbia, and a number of cities. The situation is similar in other countries; even worldwide treaties have been proposed. Specific examples:

(a) EU regulations (Browne 2024): Identifies four categories of risks: *Unacceptable, High, Limited,* and *Minimal or none*, with corresponding regulations. Includes a long list of applications (use cases) and the category each belongs in, plus a long list of regulations, including facilitating harmed parties instituting civil actions against wrongdoers. This is probably the only comprehensive policy to become law; its first provisions are in effect now, and others phase in through 2027. *April 2021*.

(b) A large number of AI-related bills have been introduced in the 118th Congress as of 31 May 2024; the Brennan Center lists about 95 (Ayoub 2024) despite omitting some that don't satisfy its definition of "AI related". Just between May and September 2023, those introduced include the Block Nuclear Launch by Autonomous Artificial Intelligence Act of 2023 (S. 1394, H.R. 2894), the REAL Political Advertisements Act (S. 1596, H.R. 3044), the AI Labeling Act of 2023 (S. 2691), the CREATE AI bill (S. 2714), Protect Elections from Deceptive AI bill (S. 2770), the AI Disclosure Act of 2023 (H.R. 3831), and the DEEPFAKES Accountability Act (H.R. 5586). As of this writing, the DEFIANCE Act (S. 3696) has passed the Senate, and a few others have passed out of committee, for instance, the Preparing Election Administrators for AI Act, S. 3897. Otherwise none of the bills introduced in the 118th Congress has gone anywhere yet.

(c) Transparency: consumers must know when they're experiencing or interacting with "fabricated media". (This plus having recourse to interact with a real person if they want is IBM's "principle" 3.) For example, a new state law on use of digitally altered media in elections (Indiana Public Law 81-2024; see 2024 Indiana Election Legislation Summary n.d.) requires that such media, whether printed, audio, or video, include a disclaimer that compiles with specific requirements. (A detailed definition of fabricated media appears in Indiana Public Law 81-2024.) Similarly, in July 2023, the nonprofit group Public Citizen petitioned the Federal Election Commission (FEC) to clarify that the existing law against "fraudulent misrepresentation" applies to deceptive AI campaign communications (Public Citizen 2023). (The FEC had not acted on the proposal by July 2024, so the FCC said they would require political advertisers to disclose their use of AI in broadcast television and radio ads. However, in August 2024, the FEC chair said that his agency won't do it and threatened to sue the FCC if *they* proceed, on the grounds that the FEC has exclusive jurisdiction over such regulations (Swenson 2024)!)

(d) Require AI companies to satisfy governments that their products are safe, at least for higher-risk use cases. (This is IBM's principle 4.) This requirement might apply only to "frontier models", those pushing the limits of technology. President Biden's 30 October 2023 executive order on AI requires companies to notify the government when they are developing models whose training involves more than a certain total number of mathematical operations, a computing cluster with more than a certain internal bandwidth, or a certain number of mathematical operations per second. The firms will then have to conduct certain safety tests on these models and share the results with the U.S. government (The White House 2023). Similarly, California's SB-1047 would require developers to implement "reasonable safeguards" against serious risks, such as mass casualties or large-scale cyberattacks, and to implement a "kill switch" (open-source models are exempted once they're out of the developer's control). Much of its requirements affect only very large models with training runs of the same size as President Biden's executive order. The most common testing methodology is "red teaming"; for particularly high-risk situations, it could be done with the AI system in a "sandbox", that is, completely isolated from other systems. A very different and important (though neglected) approach is via software security methods like architectural risk analysis (BIML, 2024).

(e) Two proposed U.S. regulations to control foreign actors, etc. (Stewart and Byrd 2023).
  1. Social media companies must either label all accounts as (1) belonging to someone verified with strong identification to be a U.S. citizen, (2) belonging to an entity verified as a specific corporation, or (3) "other"; or they must offer account owners the option of being verified to be a U.S. citizen. Verifying the owner's status *in person* might be required.
  2. Social media companies may not give users content suggestions generated via A+AI.

(f) Social media companies must wait for verification of facts or provenance before disseminating posts very widely (Giordano et al. 2023).

(g) Social media companies must "use well-designed interstitials when users engage with things like old articles or state-controlled media; utilize virality circuit breakers to automatically flag fast-spreading posts and trigger a brief halt on algorithmic amplification; restrict rampant resharing during election season by removing simple share buttons on posts after multiple levels of sharing; and implement clear strike systems to deter repeat offenses, curtail the outsized impact of malign actors, and better inform users." (Accountable Tech 2023)

(h) Bans on use of A+AI "fabricated media" such as a February 2024 ban approved by the Federal Communications Commission (FCC) on the use of artificial voices in robocalls.

(j) A new regulatory agency, like the U.S. Nuclear Regulatory Commission or FDA (OpenAI and many others), preferably with an equivalent of clinical trials in medicine to evaluate safety. (Harari 2023)

(k) A requirement for "algorithm audits" of systems used in employers' hiring processes to avoid bias (Gerchick & Akselrod 2023). This requirement could obviously be extended to other sensitive use cases.

(l) Software developers and distributors must "prevent their audio and visual products from creating harmful deepfakes, and be held liable if their preventive measures are too easily circumvented." (Disrupting the Deepfake Supply Chain 2024)

(m) The UN Secretary-General and the President of the International Committee of Red Cross have jointly called for countries to establish new prohibitions and restrictions on autonomous weapon systems ("UN Secretary-General, President of International Committee of Red Cross Jointly Call for States to Establish New Prohibitions, Restrictions on Autonomous Weapon Systems" 2023).

**P4. Licensing regulations.** Two examples:

(a) A U.S. licensing agency has been proposed (OpenAI and many others).

(b) A U.S. AI Safety Institute under the National Institute of Standards and Technology has been created to develop standards, tools, and tests. This is mandated by President Biden's "Executive Order on Safe, Secure, and Trustworthy AI" (The White House 2023).

**P5. Pause or freeze in development.** Specific proposals:

(a) 6-month worldwide pause in training of AI systems more powerful than GPT-4 (Future of Life Institute 2023)

(b) Indefinite worldwide freeze in development (Yudkowsky 2023).

**P6. International collaboration,** perhaps via a watchdog agency like the International Atomic Energy Agency or the Intergovernmental Panel on Climate Change (UN Secretary General Guterres, June 2023). Such an agency might have some enforcement powers like the former, or it might simply disseminate information, like the latter. One specific proposal is worth mentioning: An independent scientific body to develop "Living Guidelines" for testing and certifying generative AI (Bockting et al. 2023).

**P7. Publicize threats.** The point is primarily to increase the likelihood that citizens will be on their guard; it should also stimulate thinking about countermeasures. A campaign to publicize threats could be directed to a specific "audience" within the general population. Election workers are a particularly important group. An example: In December 2023, the Arizona Secretary of State hosted a training exercise for election workers from around the state involving the kinds of AI-powered attacks that they might face during the 2024 election cycle (Harris, Norden, et al. 2024). The "Preparing Election Administrators for AI" bill (S. 3897, H.R. 8353) would direct the Election Assistance Commission to create a report for election offices about relevant risks of AI.

## *(c) What protective measures are likely to be effective?*

With the ideas of Section IV(a) in mind, it's clear that many of the proposals listed above are impractical and/or unlikely to be very effective. One example is **Category P5**, a **worldwide pause** or even an **indefinite freeze** in research on and development of AI. Eliezer Yudkowsky, the author and best-known proponent of the latter (**proposal P5b**), argues that even nuclear war is preferable to allowing unfettered work on AI to continue (Yudkowsky 2023). This is quite difficult to believe, but even if he's right, world leaders are very far from taking the danger of AI seriously enough to support an indefinite freeze. **Category P1**, **wrappers**, is a second example: they may work well in specific applications where what prompts users are allowed to give can be severely restricted, but in normal situations where users can ask anything, even the built-in wrappers of LLMs have been defeated over and over (Murgia n.d.). A third example: **requiring social-media companies to verify facts or provenance** (**P3f**), which probably violates the First Amendment. As Section IV(a) points out, its practicality is also questionable. **Requiring those companies to use certain controls** (**P3g**) brings up the same objections but considerably less strongly, so that might or might not be workable.

It has been argued that **licensing requirements (P4)** would benefit only "incumbents" and would hurt newcomers (Kapoor and Narayanan 2023). In addition, courts have held that software is a form of speech, suggesting that requiring licenses for software written by Americans is a violation of the First Amendment. For both reasons licensing AI systems is unlikely to work.

Finally, **watermarking of A+AI output** (**P3c**) appears at first to be yet another hopeless approach. Bad actors will undoubtedly look for ways to remove watermarks, and they may well succeed even if the watermarks are *maximally indelible*. If they succeed, of course society will look for another way to do the watermarking. But experts have expressed doubts as to whether it's possible to stay ahead in this race (Knibbs 2023), and the history of attempts to watermark at least one medium—digital audio—is not encouraging (Wikipedia 2024c).

But this does *not* mean requiring watermarking—more generally, provenance information—is hopeless, and it's so valuable it's worth fighting for, so to speak. Watermarking could be made reasonably effective by the means the Digital Millennium Copyright Act used to protect weakly encrypted digital audio recordings, namely making it a crime to circumvent the encryption (Wikipedia 2024a). The equivalent in this case would be removing the watermark, or using a system—perhaps one built by the adversary from open source—that does not add watermarks. This would not protect against determined and well-resourced adversaries, but it would stop less serious troublemakers (groups 2 and 3 from Section IV(a), and probably the vast majority) so authorities could concentrate on the others. It's desirable to require both a prominent notice (which would make the document's provenance obvious but would be relatively easy to remove) and a watermark (which would probably not be detectable except by computer and would be much more difficult to remove). It's also desirable that the watermark be easily detectable by computer to facilitate automating responses to incoming documents. If watermarking is mandated, bad actors are likely to try to confuse citizens by adding watermarks to documents *not* produced or modified by A+AI, so consideration should be given to outlawing that behavior.

Note that how provenance information is added and how it's shown is necessarily medium-specific. It's usually thought of in terms of images, but similar considerations apply to audio, video, and even text—though the difficulty of doing so varies greatly; it's especially hard to see how text could be watermarked effectively.

Social media platforms have a great deal of control over how and how widely user postings are disseminated. One idea for reducing harm is via regulations that **target the dissemination via social** **media platforms of AI-generated content**, not the generation of content by AI. Menczer and colleagues (Menczer et al. 2023) suggest imposing requirements on content based on its reach (**P3f**). In particular, before a large number of people can be exposed to some claim, we could require its creator to prove its factuality or provenance. Political challenges to the necessary rules abound here. Establishing rules as to what claims would need to be backed up would probably be a major challenge. And, as Section IV(a) above says, agreeing on rules regarding what would constitute proof is probably hopeless; agreeing on rules for provenance seems much more doable. In any case, psychologist and misinformation researcher Jon Roozenbeek (Roozenbeek et al. 2020) argues that "prebunking"—preemptively warning and exposing people to weakened doses of misinformation, analogous to vaccines—is more effective than correcting after the fact. Using prebunking instead of refusing to disseminate the information at all weakens objections to the entire process.

A completely different and still more promising approach has been proposed by Accountable Tech (**P3g**), namely four "'soft interventions' that introduce targeted friction and context to mitigate harm" (Accountable Tech 2023). Note however that any of this might require weakening Section 230 of the Communications Decency Act, another political challenge.

Another promising idea is **a crash research project to mitigate the threats** with one or more of a variety of goals (**P2a**). Perhaps the most important in the near term is making LLMs and other AIs more trustworthy, reliable (in the sense of avoiding "hallucinations"), and explainable. This is likely to be very difficult with the currently-ubiquitous neural net; it might be best achieved with a completely different architecture. On a larger time scale, alignment of AI agents—AIs that actually have their own goals—is likely to be even more difficult, but without it, AI-caused catastrophes are easy to foresee. Note that AI agents and even teams of agents already exist (Ng 2024). Again, alignment here means having goals compatible with the wellbeing of humanity. Even specifying what goals *are* compatible is very difficult; deciding probably requires input from the experts in this area, namely philosophers who specialize in the burgeoning field of AI ethics. Other worthwhile goals for the crash project include improved watermarking, and improved detection of AI-generated content, especially deepfakes.

With respect to alignment, building a frontier AI with goals including self-preservation, putting it an air-gapped "sandbox", and trying to destroy it might show that it's capable of behavior *not* compatible with human wellbeing. (A science-fiction novel, *The Two Faces of Tomorrow* (Hogan, 1979) goes into great detail about what might happen in such a situation if it was carried out in a space station. Despite the book's age, every major idea and most details ring true today.) And a strong case can be made that making the "black boxes" of current frontier LLMs trustworthy requires applying software security methods such as architectural risk analysis to them (McGraw et al. 2024).

Hammond (2023) comments that "Policymakers cannot afford a drawn-out interagency process or notice and comment period to prepare for what's coming. On the contrary, making the most of AI's tremendous upside while heading off catastrophe will require our government to stop taking a backseat role and act with a nimbleness not seen in generations. Hence the need for a new Manhattan Project." The AI project would be similar to the original Manhattan Project in the sense of urgency behind it and the scale: to have much impact, it would need to employ many thousands, perhaps tens of thousands, of researchers. For comparison, at its peak the Manhattan Project employed some 40,500 people other than construction workers. But our problem is very different from "build an atomic bomb as soon as possible" in that what is needed to reduce the threats of AI, especially beyond the near future, is simply not that clear. It's even been argued that a crash project isn't a good idea because "ten thousand researchers have ten thousand different ideas on what [AI safety] means and how to achieve it" (Kaushik and Korda 2023). There's some truth to this. It's not a fatal objection, but it suggests the project should start with a relatively small team focused on clarifying what should be done (including philosophers to study alignment), and then grow at a rapid pace.

It's important for such a project to be under international control, for example via the nascent international network of AI safety institutes (NIST 2024). The White House announced in November 2023 the creation of the U.S. Artificial Intelligence Safety Institute (USAISI), housed within the National Institute of Standards and Technology (NIST). But USAISI probably wouldn't be an appropriate home for the project: aside from being tied to one country, the remit of NIST— USAISI's parent organization— is primarily setting standards.

How large should an AI-safety crash project really be? Menczer et al. (2023) argue that "research investments aimed at safeguarding humanity…should be comparable to the investments that are leading to current advances in generative AI". Along the same lines, Bengio et al. (2023) urge "major tech companies and public funders to allocate at least one-third of their AI R&D budget to ensuring safety and ethical use." It's been estimated that currently one researcher is employed working on safety for every 30 working on other aspects of AI, so by these standards, the world needs at least 15 times as many safety/ethics researchers as we have (1:30 vs. 15:30).  However, while current estimates of the numbers appear to be nonexistent, based on recent estimates (McAleese 2022), it seems likely that there are about 500 AI safety researchers and somewhere between 50,000 and 100,000 other AI researchers. That's a ratio of one to 100 at best. (Mustafa Suleyman, co-founder of Google DeepMind, suggests an even more radical increase is needed: he has written that we should have hundreds of thousands of researchers working on safety. But it's difficult to take the idea of *that* large a number of researchers seriously.)

**International collaboration** (**P6**) to whatever extent is possible will undoubtedly help. Robert Wright has argued that "AI has become dangerous. So it should be central to foreign policy." (Wright 2023) "Dangerous" is an understatement; otherwise this makes sense. In May, ten countries and the European Union (EU) signed the "Seoul Statement of Intent toward International Cooperation on AI Safety Science," establishing "an international network of AI safety institutes. This agreement builds on measures that several of its signatories have taken on AI safety since the Bletchley Park summit in November 2023… The Statement of Intent also builds on existing bilateral agreements." (Lang 2024) This a giant step in the right direction, but another step or two are needed, perhaps via a watchdog agency like the International Atomic Energy Agency or the Intergovernmental Panel on Climate Change (UN Secretary General Guterres, June 2023). Such an agency might have some enforcement powers, like the former, or it might simply disseminate information, like the latter. As always, reaching agreement on enforcement is likely to be very difficult. Bockting et al. (2023) urge the creation of "An independent scientific body to develop "Living Guidelines" for testing and certifying generative AI. The idea of "living" guidelines, i.e., guidelines that can evolve as needed, makes a great deal of sense, as does the idea of an "an independent scientific body" being in charge.

One more idea needs discussion, particularly since to the author's knowledge it's completely new. This is **requiring social-media companies to offer verification that account owners are U.S. citizens,** with verification requiring the **presentation of strong identification** (**P3e**). ("Strong identification" here means identification that a person is who they claim to be and that would be very difficult to fake. For instance, a passport would probably qualify; a student ID card would not, even with a photo.) Implementing that would require significant effort by social-media companies, but it would have overwhelming benefits. First and foremost, a good case can be made that in recent years American politics has been badly distorted by foreign interests, and/or by citizens *believing* it's been badly distorted by foreign interests. With strong verification, we won't mistakenly give First Amendment protection to, say, Russian agents. Second, strong verification would give legitimate posts from verified citizens more weight (which in turn would encourage citizens to voluntarily go through the verification process). And third, if the companies were required to keep track of who the owner of each verified account is, it would support non-repudiation, i.e., making it much harder for people to claim convincingly that they didn't post something they actually did post. (Disclosing the identity of an account's owner without their permission should probably require a court order.)

The recent appearance of Sam Altman's Worldcoin is helpful here. A Worldcoin white paper ("Proof of Personhood: What It Is and Why It's Needed." 2023) talks about *Proof Of Personhood* in detail; it even mentions applications of the technology to combating AI-generated disinformation. The identification having been presented *in person* might be an option that would be reflected in the verification notice shown to viewers of the account. It would likely face substantial pushback, so it should not be a requirement. Worldcoin makes an in-person option considerably more credible, since Worldcoin itself requires in-person verification of someone's identity, and for practical applications, it requires systems to do that securely and on a large scale. As of December 2023, almost 6 million people had verified their World ID with an Orb, Worldcoin's biometric imaging device. (In addition, there's now an organization, humanID , devoted to what they call "the fundamental rule of democracies: one human, one voice" ("Anonymous and Quick, Only for Humans." n.d.), but they do it in an insecure way by associating each person with a telephone number, and they don't require in-person verification, which is essential to guarantee each ID represents a unique person.)

Finally, it's not clear that *any* set of actions by governments can reduce the threats to an acceptable level unless the citizenry is on the alert for them. Therefore **major educational campaigns to raise awareness of the threats of AI (P7)** are absolutely essential. The table below relating threats and countermeasures shows why: "Publicize threats" is the only countermeasure with much promise for many of the threats. A side effect of such a campaign aimed at citizens at large would be stimulating a debate on the role of A+AI in society, thereby assisting in the search for creative ways to manage threats. But why the plural "campaigns"? Because a specialized effort is needed for at least one specific group of citizens, namely election workers. First, these days they face unique threats, e.g., attempts to overwhelm their offices with huge numbers of bad-faith requests—one is mass challenges to voter registration—that they're required to take seriously. And second, it's vitally important to minimize the spread of election-related misinformation, potentially on a very short timescale (Harris, Norden et al. 2024). An effective campaign for election workers should not be too difficult if only because there aren't that many election offices; an effective campaign aimed at the public is likely to be quite difficult, but there's no substitute for it.

## V. Legislation and/or regulations to enact now

Speaking to the danger of future AIs, Yuval Noah Harari has written (Harari 2023):

> Won't slowing down public deployments of AI cause democracies to lag behind more ruthless authoritarian regimes? Just the opposite. Unregulated AI deployments would create social chaos, which would benefit autocrats and ruin democracies.

> We should put a halt to the irresponsible deployment of AI tools in the public sphere, and regulate AI before it regulates us. And the first regulation I would suggest is to make it mandatory for AI to disclose that it is an AI. If I am having a conversation with someone, and I cannot tell whether it is a human or an AI—that's the end of democracy.

How AI will develop in the future is unclear in many ways. As a result, AI researcher Jack Clark's argument is equally important (Clark 2024):

> I've come to believe that in policy "a little goes a long way" – it's far better to have a couple of ideas you think are robustly good in all futures and advocate for those than make a confident bet on ideas custom-designed for one specific future.

Of the current ideas for making A+AI safer, those that should be put into legislation or regulations as soon as possible include:

- Require all social media platforms accessible within the U.S. to offer users verification that their accounts are owned by U.S. citizens, and to prominently display every account's verification status on everything associated with the account. Citizenship must be verified by presentation of strong identification, perhaps with an in-person option that would be displayed.

- Establish regulations on A+AI in three regards:

    o Harari's "first regulation": All products created or significantly modified with A+AI—not just generative AI—must be clearly labeled as such. This applies to AIs that converse with humans, even for mundane purposes like tech support, as well as to all pre-generated media. The details necessarily vary with the medium, but for media for which it makes sense both a prominent notice (to make the document's provenance obvious, though it would be relatively easy to remove) and a watermark (which would probably not be detectable except by computer and would be much more difficult to remove) are required. There are stiff penalties for violations. (Omitting a label intentionally for a nefarious purpose should probably also be a serious crime.)

    o Generative AI may not be used to create a likeness of a minor and may not be used to create a likeness of an adult without the advance permission of that adult.

    o Creators of generative AI software must disclose all materials used to train their software and must compensate the creators of any protected material (copyright, trade secret, trademark, or other) used in the training of the AI.

- Fund a crash project of research on mitigating the threats, with one or more of a variety of goals. Most important in the near term: making LLMs and other AIs more trustworthy, reliable (in the sense of avoiding "hallucinations"), and explainable. On a larger time scale, alignment of AI agents—AIs that actually have their own goals—is vital. Other worthwhile goals include improved watermarking of all media, and improved detection of AI-generated content, especially deepfakes. The project should be under international control, for example via the nascent international network of AI safety institutes.

- Fund large-scale educational campaigns to raise awareness of the threats among election workers and among citizens in general.

Legislative and regulatory actions on new and newly discovered threats to society will naturally evolve over time. But all of the above ideas could be implemented today and seem clearly to be constitutional, and they are likely to ameliorate quite a bit of the negative effects of A+AI that we can already see. In addition, none requires a definition of AI or judging the truth of claims. And, while AI as a whole is very difficult to define, "generative AI" is not (see Section IV(a)).

Knibbs, Kate. 2023. "Researchers Tested AI Watermarks—and Broke All of Them." Wired, October 3, 2023. https://www.wired.com/story/artificial-intelligence-watermarking-issues/.

{Lakera]. "The World's Most Advanced AI Security Platform." n.d. Accessed June 3, 2024. https://www.lakera.ai/.

Lang, Courtney. 2024. "Advancing AI safety requires international collaboration. Here's what should happen next." https://www.atlanticcouncil.org/blogs/new-atlanticist/advancing-ai-safety-requires-international-collaboration-heres-what-should-happen-next/.

Lee, Timothy B. 2023. "Joe Biden's Ambitious Plan to Regulate AI, Explained." Understanding AI. October 31, 2023. https://www.understandingai.org/p/joe-bidens-ambitious-plan-to-regulate?ref=wakeuptopolitics.com.

Lenat, Doug, and Gary Marcus. 2023. "Getting from Generative AI to Trustworthy AI: What LLMs Might Learn from Cyc." arXiv [cs.LG]. arXiv. http://arxiv.org/abs/2308.04445.

Lendon, Brad, & Nectar Gan. 2024. "China's military shows off rifle-toting robot dogs." https://www.cnn.com/2024/05/28/china/china-military-rifle-toting-robot-dogs-intl-hnk-ml/index.html.

Marinos, Alexandros, Gary Marcus, & Judea Pearl. 2023. [A Manhattan Project for AI.] March 31, 2023. https://x.com/yudapearl/status/1641978456513867776.

Marcus, Gary. 2017. "Artificial Intelligence Is Stuck. Here's How to Move It Forward." The New York Times, July 29, 2017. https://www.nytimes.com/2017/07/29/opinion/sunday/artificial-intelligence-is-stuck-heres-how-to-move-it-forward.html.

———. 2023. "Things are about to get a lot worse for Generative AI: A full spectrum of infringement". Marcus on AI, December 29, 2023. https://garymarcus.substack.com/p/things-are-about-to-get-a-lot-worse.

———. 2024. "'When looked at carefully, OpenAI's new study on GPT-4 and bioweapons is deeply worrisome." Marcus on AI. February 4, 2024. https://garymarcus.substack.com/p/when-looked-at-carefully-openais.

McAleese, Stephen. 2022. "Estimating the Current and Future Number of AI Safety Researchers." Accessed June 3, 2024. https://forum.effectivealtruism.org/posts/3gmkrj3khJHndYGNe/estimating-the-current-and-future-number-of-ai-safety.

McGraw, Gary, Harold Figueroa, Katie McMahon, and Richie Bonett. 2024. "An Architectural Risk Analysis of Large Language Models." Berryville Institute of Machine Learning. https://berryvilleiml.com/results/. https://berryvilleiml.com/docs/BIML-LLM24.pdf.

Menczer, Filippo, David Crandall, Yong-Yeol Ahn, and Apu Kapadia. 2023. "Addressing the Harms of AI-Generated Inauthentic Content." Nature Machine Intelligence 5 (7): 679–80. https://doi.org/10.1038/s42256-023-00690-w.

Mims, Christopher. 2024. "AI Doesn't Kill Jobs? Tell That to Freelancers." Wall Street Journal, June, 21, 2024. https://www.wsj.com/tech/ai/ai-replace-freelance-jobs-51807bc7.

Mizokami, Kyle. 2021. "Autonomous Drones Have Attacked Humans. This Is a Turning Point." Popular Mechanics. June 1, 2021. https://www.popularmechanics.com/military/weapons/a36559508/drones-autonomously-attacked-humans-libya-united-nations-report/.

Murgia, Madhumita. n.d. "Broken 'guardrails' for AI Systems Lead to Push for New Safety Measures." Accessed June 3, 2024. https://www.ft.com/content/f23e59a2-4cad-43a5-aed9-3aea6426d0f2

Ng, Andrew. 2024. "Agentic Design Patterns Part 1: Four AI agent strategies that improve GPT-4 and GPT-3.5 performance". https://www.deeplearning.ai/the-batch/how-agents-can-improve-llm-performance/.

NIST. 2024. "U.S. Secretary of Commerce Gina Raimondo Releases Strategic Vision on AI Safety, Announces Plan for Global Cooperation Among AI Safety Institutes." May 21, 2024. https://www.nist.gov/news-events/news/2024/05/us-secretary-commerce-gina-raimondo-releases-strategic-vision-ai-safety.

Panas, Joshua. 2023. "Mom warns of AI scam after receiving call claiming child was kidnapped". Scripps News, April 18, 2023. https://scrippsnews.com/stories/mother-falsely-told-her-daughter-was-kidnapped-in-ai-scam/.

"Pause Giant AI Experiments: An Open Letter." 2023. Future of Life Institute. March 22, 2023. https://futureoflife.org/open-letter/pause-giant-ai-experiments/.
23

# VII. Acknowledgements

Craig Stewart went far beyond the call of duty in helping me prepare this document; in addition, the important idea of social media verification of account owners' citizenship was his. Etienne Barnard made major contributions to early versions. Dave Bender, Josh Cynamon, Douglas Hofstadter, Jeff Marks, Bruce Morlan, David Moser, and Scott Pell made thoughtful comments, and David Alpher and Kevin Smith did last-minute proofreading on very little notice. Hal Turner's support in making contacts greatly increased the chances of this work ever making a difference in the world. My deepest thanks to all.

# Author Biography

Don Byrd received his Ph.D. in Computer Science from Indiana University in 1984; his dissertation supervisor was the well-known cognitive scientist Douglas Hofstadter. Byrd is noted for his contributions to music information retrieval—an AI-related field he helped to found—and music information systems. He has also worked both inside and outside academia on text information retrieval, information visualization, user-experience design, and math education, among other things. His work outside academia was for three high-tech startups, including one of his own. Byrd is the author of the open source music notation editor Nightingale. Now retired, he spends some time on music, but more working with Braver Angels (https://braverangels.org), a grassroots national organization dedicated to reducing the toxic polarization of American politics. He is one of Braver Angels' state coordinators for Indiana, and he co-chairs a Braver Angels "task force" to combat AI-driven polarization.